\documentclass[aps,pra,superscriptaddress,amsmath,amsfonts,amssymb,floatfix,nofootinbib]{revtex4}
\usepackage{enumerate}
\usepackage{mathtools}
\usepackage{amsmath}
\usepackage{graphicx}
\usepackage{epsfig}
\usepackage{sidecap}
\usepackage{hyperref,stackrel}
\usepackage[normalem]{ulem}
\usepackage{color}
\usepackage{enumerate}
\usepackage{bm}
\usepackage[utf8]{inputenc}
\usepackage{amsmath}
\usepackage{amsfonts}
\usepackage{amssymb}
\usepackage{graphicx}
\usepackage{enumitem}

\begin{document}
\title{Attainable and usable  coherence in X states over Markovian and non-Markovian channels}
\author{Sandeep Mishra} \thanks{sandeep.mtec@gmail.com }
\affiliation{Jaypee Institute of Information Technology, A-10, Sector-62, Noida, UP-201309, India}

\author{Kishore Thapliyal} \thanks{kishore.thapliyal@upol.cz}
\affiliation{Joint Laboratory of Optics of
Palack\'{y} University and Institute of Physics of CAS, Faculty of Science,
Palack\'{y} University, 17. listopadu 12, 771 46 Olomouc, Czech Republic}

\author{Anirban Pathak} \thanks{anirban.pathak@gmail.com}
\affiliation{Jaypee Institute of Information Technology, A-10, Sector-62, Noida, UP-201309, India}

\begin{abstract}
The relations between the resource theoretic measures of quantum coherence are rigorously investigated for various Markovian and non-Markovian channels for  the two-qubit $X$ states with specific attention to the maximum and minimum attainable coherence and usefulness of these states in performing quantum teleportation in noisy environment. The investigation has revealed that under both dephasing and dissipative type noises the maximally entangled mixed states and Werner states lose their form and usefulness. However, maximally non-local mixed states (MNMSs) lose their identity in dissipative noise only. Thus, MNMSs are established to be useful in teleporting a qubit with fidelity greater than the classical limit in the presence of dephasing noise. MNMSs also remain useful for device independent quantum key distribution in this case as they still violate Bell's inequality.  In the presence of noise, coherence measured by relative entropy of coherence is found to fall faster than the same measured using $l_1$ norm of coherence.  Further, information back-flow from the environment to the system is observed over non-Markovian channels which leads to revival in coherence. Additionally, sequential interaction of two qubits with the same environment is found to result in correlated noise on both qubits, and coherence is observed to be frozen in this case under dephasing channel. Under the effect of Markovian and non-Markovian dephasing channels studied here, we observed that MNMSs have maximum relative coherence, i.e., they have the maximum amount of $l_1$ norm of coherence among the states with the same amount of relative entropy of coherence. However, this feature is not visible in any $X$ state evolving over dissipative channels.  

\end{abstract}

\maketitle

\section{Introduction} \label{Intro}

Coherence has been known to us for a few centuries in the context of classical optics, but the extension of this idea to quantum systems has added an altogether new dimension \cite{Mandel,glaub,sudar}. Quantum coherence which originates from the wave function description of quantum systems is  considered as one of the most important characteristic features of any quantum system. Other than the physical processes \cite{coh_rev2,coh_rev1}, recently it has been recognized that quantum coherence plays a very important role in many biological processes, too, such as photosynthesis \cite{bio1}, avian compass in migratory birds \cite{bio2} and quantum biology \cite{bio3}. Even though quantum coherence has been known to exist  for long, its quantification is proposed only recently. In 2014, Baumgratz et al. \cite{plenio} provided a quantitative framework to identify, characterize and manipulate coherence present in the quantum systems and introduced two measures known as relative entropy of coherence and $l_1$ norm of coherence. 
 The $l_1$ norm of coherence \cite{plenio} of quantum state $\rho$ is given by
\begin{equation}\label{l1}
C_{l_1}(\rho)= \sum _{i,j\forall i\neq j}|\rho_{i,j}|.
\end{equation}   
The relative entropy of coherence \cite{plenio} present in a quantum state represented by the density matrix $\rho$ is defined as 
\begin{equation}\label{rel_ent}
C_{\rm{} rel}(\rho)= S(\rho_{\rm{} diag})- S(\rho),
\end{equation}
where $S(\rho )$ is the von Neumann entropy of $\rho$, and $\rho_{\rm{} diag} $ denotes the state obtained from $\rho $ by removing all the off-diagonal elements of $\rho$. 
Both these quantities are basis dependent. These two coherence measures are individually considered as good measures of coherence even though both have vastly different physical interpretations. To be precise, relative entropy of coherence represents  the optimal rate of the distilled maximally coherent states that can be produced by incoherent operations in the asymptotic limit of many copies of $\rho$ \cite{winter}. The $l_1$ norm of coherence has been reported to be a good quantifier of the wave nature of a quanton in a multi-slit interference set up and this can be experimentally measured \cite{bera, sandeep1, sandeep2}.  The study of resource theory of coherence and its applications in various fields has now become a very active area of research \cite{coh_rev2,coh_rev1}. 

Coherence of various finite and infinite dimensional quantum states is reported in the recent past (\cite{mishra_x,optical_coh,coh-res} and references therein). A specific quantum state of interest is $X$ state introduced by Yu and Eberly \cite{xstate1, xstate2} in a study highlighting the finite-time disentanglement of two-qubits due to spontaneous emission resulting in entanglement sudden death \cite{esd}. These are known as $X$ states due to the positions of the non-zero elements of $\rho_X$ in the computational basis $ \{ \vert 00 \rangle$, $ \vert 01 \rangle$, $ \vert 10 \rangle$,$ \vert 11 \rangle \}$ resemble the shape of the letter $X$ as \cite{xstate1, xstate2}
\begin{equation} \label{x_state}
\rho_{X} = \left(
     \begin{array}{cccc}
      \rho_{11} & 0 & 0& \rho_{14}  \\
       0&\rho_{22} &\rho_{23}& 0 \\
       0 & \rho^{*}_{23} &\rho_{33}& 0 \\
      \rho^{*}_{14} & 0 & 0 & \rho_{44} \\
     \end{array}
   \right).
\end{equation}
For $\rho_{X} $ to represent a physical state, we must have $ \underset{i}{\sum}\rho_{ii}=1 $, $ \rho_{22}\rho_{33}\geq |\rho_{23}|^2$, and $ \rho_{11}\rho_{44}\geq |\rho_{14}|^2$ \cite{x_dynamics}.  Interestingly, the $X$ states can be separable or entangled depending upon the values of parameters describing them. It is known that $X$ states are entangled \cite{x_dynamics} if and only if either $\rho_{22}\rho_{33} < |\rho_{14}|^{2}$ or $\rho_{11}\rho_{44} < |\rho_{23}|^{2}$, and the amount of entanglement, as measured by concurrence, is given by $\max \left\{0, 2\left(|\rho_{14}|- \sqrt{\rho_{22}\rho_{33}}\right), 2\left(|\rho_{23}|- \sqrt{\rho_{11}\rho_{44}}\right) \right\}.$ 

Since their introduction, these states have become a subject of extensive study as they contain an important class of pure and mixed states, such as maximally entangled states (like Bell states), partially entangled and quantum correlated states (like the Werner states \cite{werner}), maximally nonlocal mixed states (MNMSs) \cite{mnms}, maximally entangled mixed states (MEMSs) \cite{munro,ishizaka,verst} as well as non-entangled (separable) states. Recently, Paulo et al. \cite{xh} have proved that for every two-qubit state, there is a $X$-counterpart, i.e., a corresponding two-qubit $X$ state having the same spectrum and entanglement, as measured by concurrence, negativity or relative entropy of entanglement. This universality property of  $X$ states allows us to map every two-qubit state to an equivalent $X$ state, and hence these $X$ states constitute an important resource for quantum communication and computation. Due to the presence of the elements in the main and other diagonal only, the two-qubit $X$ states have been shown to be produced in a vast variety of  experimentally realizable systems such as optical systems \cite{xp_p1,xp_p2,xp_p3,xp_p4}, ultra cold atoms \cite{xp_uc1,xp_uc2}, and nuclear magnetic resonance \cite{xp_nmr}. More recently, a method for generation of spin-orbit $X$ states via the use of an optical setup is also proposed \cite{x2021}.

In an earlier work \cite{mishra_x}, we explored the issues related to the ordering of quantum states based on the amount of coherence as measured by any two measures of coherence for general two-qubit states with a specific attention to the study of $X$ states. Further, we identified the states lying at the boundaries for the scatter plot of coherence of the $X$ states measured by $l_1$ norm of coherence and relative entropy of coherence. In other words, we identified quantum states with maximum relative amount of coherence quantified by $l_1$ norm of coherence among the set of quantum states with the same amount of relative entropy of coherence. However, in a realistic scenario, a quantum state can  not be free from the effect of the surrounding environment \cite{nc} which usually degrades the quantum effects, such as coherence, discord, entanglement, and hence their usefulness in quantum compuation, communication and metrology \cite{pet_book}. When the system interacts weakly with the environment, then the correlation functions of the environment decays at a much shorter time scale in comparison to the decay of correlations for the system. This approximation is known as Markov approximation and the reduced dynamics of the system is defined by linear maps known as Markovian channels \cite{QC-rev,kraus,NM-rev}. Here, the evolution of system is just dependent upon the present state and does not take into account any previous state of the system. However, there may be situations where the Markovian approximation is not valid, such as strong interactions with the environment. For such a situation, the reservoir correlation time is of the same order to the decoherence time of the system. Such dynamics is described by non-Markovian channels. The dynamics of the system over these two types of channels has specific characteristics. For instance, invertible dynamical maps are P-divisible for Markovian case but not otherwise. A quantum process is Markovian if the distance between two arbitrary quantum states decreases monontonically \cite{NM-rev}. This corresponds to the loss of information from the system to the environment. In contrast, there can be a back-flow of information from the environment to the system for the non-Markovian channels.  Independently, quantum channels with memory are reported where Markovian correlated noise acts during multiple uses of a transmission channel or sequential interaction with a common environment \cite{QC-rev}. Recent studies have shown a relation between the quantum channels with memory and non-Markovian dynamics. Specifically, it is shown that the dynamics of the second subsystem due to correlated noise become non-Markovian though the first subsystem undergone a Markovian dynamics \cite{NM-mem}.

Recently, there has been a slew of papers on the study of coherence and other quantum features such as entanglement and discord under the effect of different Markovian and non-Markovian noises \cite{frozen2015,guo,liu2017,huang2017,radha2019,young2020,song2020,zhao2020,wang2019,luo2019,jiang2020,zhang2019,cai2020,naikoo1,naikoo2,naikoo2020,passos2019}. 
The conditions under which the coherence of a single qubit and Bell diagonal states can be frozen under the action of Markovian channels are obtained \cite{frozen2015}. The evolution of coherence for two-qubit states under the action of correlated noise showed that it is effective in protecting the decay of coherence \cite{guo}. Non-Markovian dynamics of a two-level system under the presence of an external driving field is also reported \cite{huang2017}. The resilient effect of quantum coherence in GHZ states under the action of  bit-flip noise is experimentally investigated \cite{zhang2019}. The relation between coherence, entanglement and discord for the $X$ states under the action of some Markovian noises has also been reported recently \cite{young2020}. Similarly, the evolution of the relative entropy of coherence for the $X$ states under the multiple application of amplitude damping noise is also studied \cite{song2020}. Jiang et al. studied the trade-off relations between the $l_1$ norm of coherence for a multipartite system \cite{jiang2020}. Along the same lines, some works have been performed to study the effect of non-Markovian noises \cite{naikoo1,naikoo2,naikoo2020,passos2019,cai2020}. These recent studies motivated us to obtain the maximum and minimum attainable coherence measured by different resource theoretic measures intending to acquire a quantitative perception of the usefulness of the $X$ states in realistic conditions. Specifically, in this work, we investigate the evolution of the coherence of the $X$ states using $l_1$ norm and relative entropy of coherence under the action of some well-known Markovian and non-Markovian channels. Further, as an example, usefulness of $X$ states in quantum teleportation over noisy channels is also investigated.

The rest of the paper is organized as follows. In Section \ref{sec2}, we study the evolution of the coherence of the $X$ states  under some of the Markovian and non-Markovian dephasing type noises. This is followed by the discussion over some dissipative type noises in Section \ref{sec3}. Subsequently, in Section \ref{sec4}, we analyze the usefulness of $X$ states as a resource for the process of teleportation under the presence of noisy environment. Finally, we conclude by summarizing all the results in Section \ref{sec5}.

\section{$X$ states under dephasing quantum channels} \label{sec2}

Evolution of a quantum system under the effect of ambient environment can be described in terms of trace preserving dynamical maps, i.e., quantum channels (\cite{QC-rev} and references therein). Specifically, a quantum system evolving under the effect of noise is described by a completely positive trace preserving (CPTP) dynamical map from an initial state $\rho$ to the final state $\rho'$ in operator sum representation \cite{kraus} as
\begin{eqnarray}\label{Kr}
\rho'= \sum _{i} E_{i}\rho E_{i}^{\dagger},\label{eq:Kr}
\end{eqnarray}
where $E_{i}$s are the Kraus operators with $\sum_{i}E_{i}^{\dagger} E_{i} = I$.
In this section, we discuss the effect of Markovian dephasing channels and subsequently non-Markovian dephasing channels on the coherence in $X$ states. 
The dephasing channels are noiseless for transmission of classical basis states while the superposition of basis states undergoes decoherence.

\subsection{$X$ states under quantum Markovian dephasing channels}

A Markovian quantum channel, i.e., memoryless channel, is an element of one-parameter semigroup of CPTP maps. In this case, the dynamics of a quantum state at time $t_0$ does not depend on the state of the system before $t_0$ and the quantum and initially the quantum system is considered to be uncorrelated with its environment.
Let us first study the evolution of $X$ states under some popular Markovian dephasing channels.

\subsubsection{Phase damping channel}

Phase damping channel models the decoherence in realistic physical situations in which there is no loss of energy while the phases die out progressively. The Kraus operators for a single qubit representing the phase damping channel \cite{nc} are given by:
\begin{equation}\label{PD-Kr}
E^{\rm PD}_0 = \left(
     \begin{array}{cc}
      1 & 0  \\
       0 & \sqrt{1-\eta_{p}}    
     \end{array}
   \right)\quad {\rm and} \quad
   E^{\rm PD}_1 = \left(
     \begin{array}{cc}
      1 & 0 \\
       0 & \sqrt{\eta_{p}}     
     \end{array}
   \right),
\end{equation}
where $\eta_{p}$ is the phase damping parameter and lies between $0$ and $1$ with $\eta_{p}=0$ representing noiseless channel. For two-qubit systems, the two-qubit Kraus operators for phase damping channel are expressed as $E^{\rm PD}_i \otimes E^{\rm PD}_j \, \forall i, j \in \{0,1 \}$.  

Let us consider the effect of phase damping channel acting on both the qubits of the $X$ states (\ref{x_state}). The resultant $X$ state after the action of phase damping channel can be written, using Eq.~(\ref{PD-Kr}) in Eq.~(\ref{Kr}), as
\begin{equation} \label{x_state_PD}
\rho_{X}^{\rm PD} = \stackrel[\substack{i,j=0\\2i+j=k-1}]{1}{\sum} \rho_{kk}  \left|ij \right\rangle\left\langle ij \right| + \zeta \left( \rho_{14} \left|00 \right\rangle\left\langle 11 \right|+\rho_{23} \left|01 \right\rangle\left\langle 10 \right| +{\rm H.c.}\right),
\end{equation} 
where $\zeta=(1-\eta_{p})$ for phase damping channel.
We can clearly observe from Eq.~(\ref{x_state_PD}) that even under the action of phase damping channel, the $X$ states retain their form and consequently all the conclusions of \cite{mishra_x} related to coherence of $X$-state would remain applicable here.  

The $l_1$ norm of coherence (\ref{l1}) $C_{l_1}(\rho)=\mathcal{M}_0(\rho)$, hereafter, for the resultant state $\rho_{X}^{\rm PD}$ under the presence of phase damping channel can be computed as
\begin{equation} \label{l1_state_PD}
\mathcal{M}_0(\rho_{X}^{\rm PD})= 2 \zeta\{ \vert \rho_{14} \vert + \vert  \rho_{23} \vert \} = \zeta\mathcal{M}_0(\rho_{X})
\end{equation}
with $\zeta=(1-\eta_{p})$, where $\mathcal{M}_0(\rho_{X})=2 \{ \vert \rho_{14} \vert + \vert  \rho_{23} \vert \}$ is the coherence in the initial $X$ states (obtained for $\eta_{p}=0$).

Similarly, the relative entropy of coherence (\ref{rel_ent})  $C_{\rm rel}(\rho)=\mathcal{M}_1(\rho)$, hereafter,  for the resultant state $\rho_{X}^{\rm PD}$ evolved over the phase damping channel can be as
\begin{equation}\label{crl_state_PD}
\mathcal{M}_1(\rho _{X}^{\rm PD})= \sum_{i} \lambda _{i}^{\rm PD} \log_{2}(\lambda_{i}^{\rm PD})- \sum_{i} \rho _{ii} \log_{2}(\rho _{ii})  , 
\end{equation}
where $\lambda _ {i}^{\rm PD} $s are the eigenvalues of the state $\rho_{X}^{\rm PD}$, and we have used the fact that $\rho _{ii}^{\rm PD}=\rho _{ii}$. 
Interestingly, $X$ states maintain their form after evolution which allows us to use the eigenvalues $
\lambda _{i}(\rho_{X})$ of an arbitrary state $\rho_{X}$ in computing $\mathcal{M}_1(\rho_{X})$ as 
\begin{equation}\label{EV}
\lambda _{i}(\rho_{X})=\left\{\frac{1}{2}\left[(\rho_{11}+\rho_{44})\pm \sqrt{(\rho_{11}-\rho_{44})^{2}+ 4|\rho_{14}|^{2}} \right],\, 
\frac{1}{2}\left[(\rho_{22}+\rho_{33})\pm \sqrt{(\rho_{22}-\rho_{33})^{2}+ 4|\rho_{23}|^{2}} \right]\right\}. 
\end{equation}
{ Notice that for $\eta_p=1$, the off-diagonal elements vanish, and thus both $\mathcal{M}_0$ and $\mathcal{M}_1$ also become zero.}

\begin{figure}[tb]
\includegraphics[width=\textwidth]{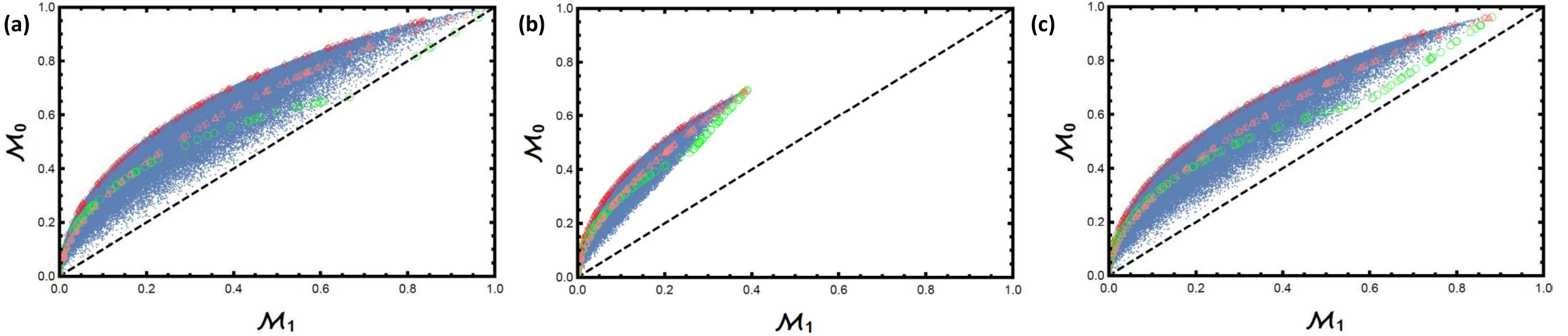}
\caption{ (Color online) { The blue points represent scatter plots for randomly prepared $X$ states evolved under phase damping noise with (a) $\eta_{p}=0 $ and (b) $\eta_{p}=0.3 $. (c) $X$ states under the action of phase damping with memory for $\eta_{p}=0.3$ and memory component  $\mu=0.9$. The red squares, green circles and pink triangles represent the same for MNMSs, MEMSs and Werner states, respectively. The black (dashed) line represents a line with slope 1.}} 
\label{fig1}
\end{figure}

To begin with, we aim to study the evolution of relative coherence of $\mathcal{M}_0$ with respect to $\mathcal{M}_1$. Specifically, in Fig. \ref{fig1} (a) we have shown 
the scatter plot of the coherence of {1 million randomly generated} $X$ states as calculated via measure $\mathcal{M}_0$ (\ref{l1}) on the $y$ axis and $\mathcal{M}_1$ (\ref{rel_ent}) on the $x$ axis. Subsequently, in Fig. \ref{fig1} (b), we present the scatter plots of the relative coherence of $X$ states evolved over phase damping channels with $\eta_{p}=0.3$. Since the $X$ states retain their form under the presence of phase damping channels, so we can observe that Rana et. al.'s \cite{rana} conjuncture that {$\mathcal{M}_0 \geq \mathcal{M}_1$} also holds true for $X$ states under the effect of phase damping noise.  We can observe that with an increase in damping parameter $\eta_p$, the maximum attainable coherence reduces with respect to both the measures of coherence. Consequently, the length of the region occupied by the scatter plot is found to reduce. Physically, this nature is expected as the enhanced noise leads to more decoherence (or reduction of coherence) and the same nature is expected to be observed under the effect of all Markovian channels. Further, we can observe that the relative coherence of $\mathcal{M}_0$ with respect to $\mathcal{M}_1$ shifts farther from the line $\mathcal{M}_0=\mathcal{M}_1$ with increasing decoherence rate.  Moreover, the spread of the scatter plots can also be observed to shrink with increasing phase damping noise which can be {considered as the} contractive behavior of CPTP map as the evolved states tend to come closer to each other \cite{NM-rev}.  Both these features suggest that the amount of coherence measured through $\mathcal{M}_1$ falls rapidly in comparison to $\mathcal{M}_0$ and the same is reflected clearly in Fig. \ref{fig1}. In what follows, we will analyze this behavior in detail, but before that we would like to highlight some other key features. Specifically, in \cite{mishra_x}, we have obtained that a class of $X$ states forms the boundary states in variation of relative coherence. Here, we verify whether those special states maintain this nature after decoherence.

For instance, we study here the evolution of the MNMSs \cite{mnms} described as 
\begin{equation}\label{mnms_state}
\rho_{\rm{} MNMS} =\frac{1}{2} \left(\left|00 \right\rangle\left\langle 00 \right|+\left|11 \right\rangle\left\langle 11 \right|+\epsilon \left\{ \left|00 \right\rangle\left\langle 11 \right| +\left|11 \right\rangle\left\langle 00 \right|  \right\}
   \right),
\end{equation}
which form a subclass of $X$ states and has a highest relative coherence of $\mathcal{M}_0$ with respect to $\mathcal{M}_1$ (as shown in Fig. \ref{fig1} (a) and also reported in \cite{mishra_x}).
For each value of $0<\epsilon\leq 1$, the state $\rho_{\rm{} MNMS}$ is a Bell diagonal state and represents the state that produces a maximal violation of the Clauser-Horne-Shimony-Holt (CHSH) inequality \cite{chsh}. Under the effect of phase damping the MNMS evolves to 
\begin{equation}\label{mnms_state_PD}
\rho_{\rm{} MNMS}^{\rm{} PD} =\frac{1}{2} \left(\left|00 \right\rangle\left\langle 00 \right|+\left|11 \right\rangle\left\langle 11 \right|+ \zeta \epsilon \left\{ \left|00 \right\rangle\left\langle 11 \right| +\left|11 \right\rangle\left\langle 00 \right|  \right\}
   \right).
\end{equation}
Interestingly, we can notice in Eq.~(\ref{mnms_state_PD}) that MNMSs retain their form even under the phase damping noise which can also be verified from Fig. \ref{fig1} (b), where we observe that the MNMSs after the effect of phase damping channel remain at the upper boundary of the scatter plots for relative coherence. This behavior can be observed for any value of phase damping parameter ($\eta_{p}$).
Another important subclass of $X$ states having enormous applications is Werner states (described as a statistical mixture of a maximally entangled state and a maximally mixed state) and can be written as 
\begin{equation}\label{werner_state}
\rho_{\rm{} W} = \frac{1+\epsilon}{4} \left(\left|00 \right\rangle\left\langle 00 \right|+\left|11 \right\rangle\left\langle 11 \right|\right)+\frac{1-\epsilon}{4}\left(\left|01 \right\rangle\left\langle 01 \right|+\left|10 \right\rangle\left\langle 10 \right|\right)+\frac{\epsilon}{2}  \left( \left|00 \right\rangle\left\langle 11 \right| +\left|11 \right\rangle\left\langle 00 \right|  \right).
\end{equation}
Separability of Werner states depends on the value of $\epsilon$. Specifically, a Werner state {is} entangled if $\epsilon > \frac{1}{3}$ and separable otherwise \cite{werner2}. Similarly, MEMSs \cite{munro,ishizaka,verst} form another subclass and can be represented as 
\begin{equation}\label{mems_state}
\rho_{\rm MEMS} =  g (\epsilon) \left(\left|00 \right\rangle\left\langle 00 \right|+\left|11 \right\rangle\left\langle 11 \right|\right)+\left(1- 2g (\epsilon)\right)\left|01 \right\rangle\left\langle 01 \right|+\frac{\epsilon}{2}  \left( \left|00 \right\rangle\left\langle 11 \right| +\left|11 \right\rangle\left\langle 00 \right|  \right),
\end{equation}
where
\begin{equation*}
g (\epsilon)= \begin{cases}
  \frac{\epsilon}{2}, & \text{if $\epsilon \geq \frac{2}{3};$ } \\
  \frac{1}{3}, & \text{$\epsilon < \frac{2}{3}$}.
\end{cases}
\end{equation*}
This class represents the quantum states for which no additional entanglement can be produced by global unitary operations. These states  are a  generalization of the class of Bell states to mixed states and are known to have the  highest degree of entanglement for a given purity of a state. We can clearly observe that unlike the case of a MNMS which transforms to another MNMS under phase damping, the Werner states and MEMSs under phase damping would not  retain their form. However, they will remain $X$ states. This is so because the last terms in the right-hand-side of Eqs.~(\ref{werner_state}) and (\ref{mems_state}), will have an additional factor of $(1-\eta_{p})$ which will reduce coherence. Consequently, usefulness of these states for information theoretic tasks will reduce in those cases where coherence is used as the essential quantum resource to perform the task.

To compare the rate of change of both the measures of coherence $\mathcal{M}_0$ and $\mathcal{M}_1$, we obtain the first derivatives of these quantities with respect to the decoherence parameter ($\eta_p$). Specifically, the rate of change of $\mathcal{M}_0$ and $\mathcal{M}_1$ can be computed as
\begin{eqnarray} \label{rate}
\frac{d \mathcal{M}_0 (\rho_{X}^{\rm PD})}{d \eta_p} &=& 2\{ \vert \rho_{14} \vert + \vert  \rho_{23} \vert \} \frac{d \zeta}{d \eta_p}= -\mathcal{M}_0(\rho_{X}), \nonumber \\  
\frac{d \mathcal{M}_1(\rho_{X}^{\rm PD})}{d \eta_p} &=& \sum_i \frac{1}{\ln(2)}\{1+\ln (\lambda_i^{\rm PD})\} \frac{d\lambda_i^{\rm PD}}{d \zeta}\frac{d \zeta}{d \eta_p}.
\end{eqnarray}

\begin{figure}[tb]
\includegraphics[width=\textwidth]{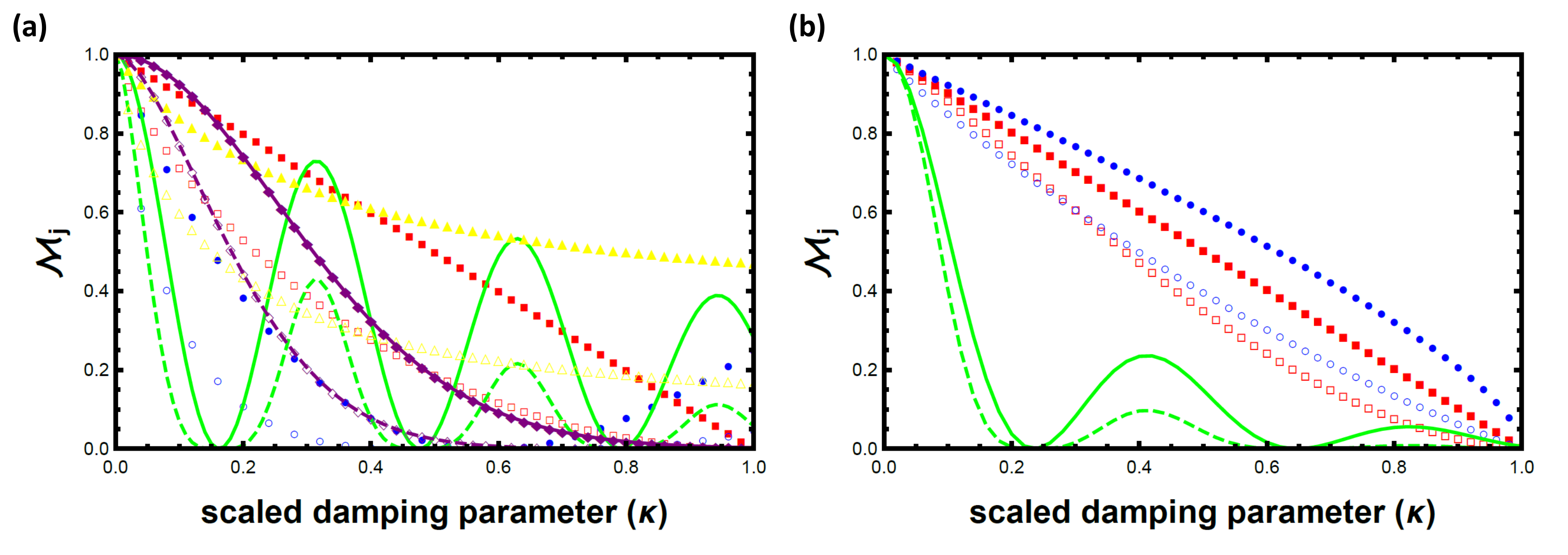}
\caption{(Color online) { Variation of  $\mathcal{M}_0$ (in solid markers) and  $\mathcal{M}_1$ (in empty/dashed markers) with respect to the scaled noise parameters ($\kappa$) for initial Bell state evolving under different (a) dephasing and  (b) dissipative type noises. In (a), results under phase damping (red squares) with $\kappa= \eta_p$, non-Markovian dephasing (blue circles) with $\alpha=1.0$ and $\kappa= \eta_p$, RTN (green lines) with $b=10\gamma$ and $\kappa=0.5 \gamma t$, PLN  (yellow triangles) with $\Gamma= 16 \gamma$ and $\kappa= \gamma t$,  OUN (purple squares with line) with $\Gamma= 16 \gamma$ and $\kappa= \gamma t$ are shown. In (b), we show results under amplitude damping (red squares) with $\kappa= \eta_a$, amplitude damping with memory (blue circles) with $\kappa= \eta_a$  and $\mu =0.5$ and non-Markovoian amplitude damping (green lines) with $ \Gamma = 0.1 \gamma$ and $\kappa=3.5 \gamma t$.}}
\label{fig2}
\end{figure}

The closed form analytic expression for the rate of change of relative entropy of coherence for MNMSs  is computed as 
\begin{eqnarray}
\frac{d \mathcal{M}_1(\rho_{\rm MNMS}^{\rm PD})}{d \eta_p} =  \frac{\epsilon}{\ln(2)} \tanh^{-1} [\epsilon (1-\eta_p)]. 
\end{eqnarray}
	A comparison of the rate of change of both the measures of coherence with respect to the decoherence parameter $\eta_p$ for the MNMSs is shown in Fig. \ref{fig2} (a). We can clearly observe that $\mathcal{M}_1$ falls at a greater rate in comparison to $\mathcal{M}_0$. Interestingly, this maintains  the validity of Rana et al.'s conjuncture \cite{rana} for all the $X$ states under the effect of phase damping. Clearly, the rate of change of $\mathcal{M}_0$ is linear while $\mathcal{M}_1$ falls faster as expected (cf. Eq. \ref{rate} and Fig. \ref{fig2}). 

Similarly, the analytic expression for the rate of change of $\mathcal{M}_1$ for the Werner states is also computed in a compact form as 
\begin{eqnarray}
\frac{d \mathcal{M}_1(\rho_{W}^{\rm PD})}{d \eta_p} =  \frac{\epsilon }{2\ln(2)} \ln \left\{\frac{1+\epsilon + 2 \epsilon(1- \eta_p)} {1+\epsilon - 2 \epsilon(1- \eta_p)}\right\}.
\end{eqnarray}
 A variation of the rate of change of $\mathcal{M}_0$ and $\mathcal{M}_1$ with respect to the  decoherence parameter $\eta_p$ for the MEMSs and Werner states is similar in nature. In the next section, we consider the effect of correlation between the two consecutive uses of the phase damping channel.

\subsubsection{Phase damping with memory}

In the case of correlated Markovian noise, the evolved state for two consecutive uses of the channel 
\cite{PD-mem} can be written as
\begin{eqnarray}\label{eq:mem}
\rho^{\prime}= (1-\mu)\sum _{i,j} E^{u}_{i,j}\rho E_{i,j}^{u\dagger} + \mu \sum _{k} E^{c}_{k,k}\rho E_{k,k}^{c\dagger} ,
\end{eqnarray}
where $E^{u}_{i,j}$ is the uncorrelated noise while $E^{c}_{k,k}$ is the correlated noise, $\rho\,\left(\rho^{\prime}\right)$ is the initial (evolved) state. { One can observe that for $\mu=0$ Eq.~(\ref{eq:mem}) reduces to Eq.~(\ref{Kr}).} { In this model, $\mu$ is known as the memory coefficient which corresponds to the classical correlation in the Markovian noise on each subsystem  interacting with the same environment sequentially.} The value of $\mu$ lies between $0$ and $1$ with $\mu=0$ corresponding to memory-less channel while $\mu=1$ representing to the complete memory channel.
 The Kraus operators for both correlated and uncorrelated noise for phase damping with memory is given in terms of Pauli operators \cite{guo} $\sigma_{0}= I$ and $\sigma_{3} = \sigma_{z}$. Specifically, the Kraus operators for uncorrelated noise are given by
\begin{equation}
E^{u}_{i,j}= \sqrt{p_{i}p_{j}} \sigma_{i} \otimes \sigma_{j}, \,(i,j = 0,3),
\end{equation}
where $p_0=1-p$ and $p_3=p$ with $0 \le p \le 1/2$. The parameter $p$ here is related to the decoherence parameter ($\eta_p$) introduced for phase damping noise as $\eta_p = 1- (1-2p)^{2}$.
Further, the Kraus operators for correlated noise is given by
\begin{equation}
E^{c}_{k,k}= \sqrt{p_{k}} \sigma_{k} \otimes \sigma_{k},\, (k = 0,3).
\end{equation}

The evolved $X$ state under the action of phase damping with memory can be calculated in the same form as in Eq.~(\ref{x_state_PD})
where $\zeta=\Omega(p,\mu) = (1-2p)^2 + 4 \mu p(1-p)=1-\eta_p+\mu \eta_p$.
The coherence measures $\mathcal{M}_0$ and $\mathcal{M}_1$ for the $X$ state evolved over the phase damping channels with memory can be obtained in the same form as Eqs.~(\ref{l1_state_PD}) and (\ref{crl_state_PD}) with $\zeta=\Omega(p,\mu)$. 

If we consider the memory component as zero (i.e., $\mu=0$), then we can deduce all the results that we have obtained in the case of uncorrelated phase damping channel (discussed in the previous subsection). We can clearly observe from Fig. \ref{fig1} (c) that the characteristic features of phase damping channel with memory are similar to that of uncorrelated phase damping channel. The presence of non-zero memory component $\mu$ plays a key role in stalling the process of decoherence. Fig. \ref{fig1} (b) shows the scatter plots of the $X$ state's coherence in the absence of memory component (i.e. $\mu=0$) for damping parameter $\eta_p=0.3$. Fig. \ref{fig1} (c) represents the same in the presence of memory with memory component $\mu=0.9$.  It can be clearly  observed that with an increase in the value of the memory component ($0 < \mu \le 1 $), the amount of the coherence can be sustained. In other words, coherence increases with $\mu$ as shown in Figs. \ref{fig1} (b) and (c).  Interestingly, in case of the $X$ states evolved over the perfectly correlated dephasing noise (i.e., $\mu=1$), we have $\Omega(p,\mu) =1$ or $\frac{d \Omega(p,\mu)}{d p} =0$ in
\begin{equation}
\frac{d \mathcal{M}_0 (\rho_{X}^{\rm PDM})}{d p}=\mathcal{M}_0(\rho_{X})\frac{d \Omega(p,\mu)}{d p}.
\end{equation}
Hence, one can clearly observe that $\mathcal{M}_0$ (and similarly $\mathcal{M}_1$) of the evolved $X$ states in this case remain unchanged with the damping parameter ($\eta_p$). Therefore, in the presence of phase damping channel with memory, the coherence of $X$ states (as well as all the parameters of the systems) gets frozen when  the memory component becomes complete ($\mu = 1$). Thus, in this situation,  the scatter plot of $\mathcal{M}_0$ and $\mathcal{M}_1$ will be same as shown in Fig. \ref{fig1} (a) irrespective of the value of decoherence parameter. Moreover, the behaviour of special classes of states and the MNMSs forming the boundary states remains valid for all values of memory component. So we can observe that the action of phase damping with memory is very much effective in preserving the coherence of the system.

In view of the recent results \cite{NM-mem} connecting the quantum channels with memory and non-Markovian dynamics, the advantage in preserving the coherence of  the system due to the classical correlation (i.e. $\mu\neq0$) can be attributed to the non-Markovian features.
In what follows, we will consider the evolution of $X$ states under some non-Markovian dephasing channels to verify this further.

\subsection{$X$ states under quantum non-Markovian dephasing channels}

Unlike the case of Markovian channels, for non-Markovian channels there is a strong coupling between system and the ambient environment which leads to back-flow of energy from the environment to the system. Let us now consider the effect of some popular non-Markovian dephasing channels on the evolution of $X$ states.

\subsubsection{Non-Markovian dephasing}

This channel is an extension of dephasing channel to non-Markovian \cite{nmd2018}. The Kraus operators are given by 
\begin{eqnarray}\label{NMD-Kr}
E_{0}^{\rm NMD}=  \sqrt{(1-\alpha p) (1-p)} \sigma_{0},  \quad {\rm and} \quad E_{1}^{\rm NMD} = \sqrt{p + \alpha p(1-\alpha p) } \sigma_{3}. 
\end{eqnarray}
Here, the parameter $\alpha$ quantifies the degree of non-Markovianity of the channel while $p$  is decoherence parameter with $0 \le {p} \le 1/2 $.  Under the effect of this noise a $X$ state evolves to the form same as in Eq.~(\ref{x_state_PD}) with $\sqrt{\zeta}=\Upsilon(p,\alpha) = 1-2p-2 \alpha p (1-p)$, which is obtained using Eq.~(\ref{NMD-Kr}) in Eq.~(\ref{Kr}). For Markovian evolution we use $\alpha=0$, while $0<\alpha\leq 1$ corresponds to the non-Markovian evolution. For $\alpha=0$ (Markovian regime) we get back all the results obtained for the action of phase damping noise for $\eta_p = 1- (1-2p)^{2}$.  For the $X$ state under the action of non-Markovian dephasing channel, $\mathcal{M}_0$ and $\mathcal{M}_1$ are obtained in the same form as in Eqs.~(\ref{l1_state_PD}) and (\ref{crl_state_PD}) with $\sqrt{\zeta}=\Upsilon(p,\alpha)$.

\begin{figure}[tb]
\includegraphics[width=\textwidth]{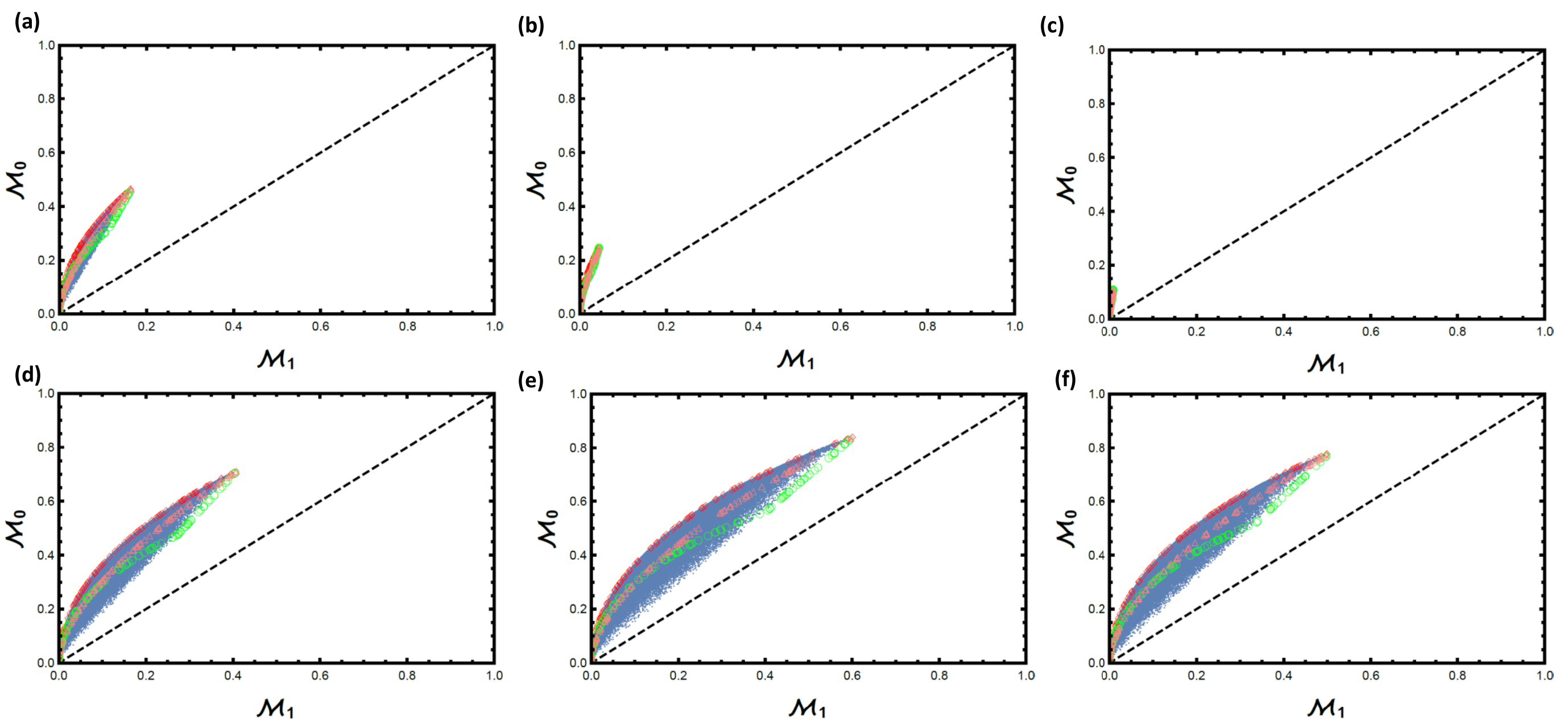}
\caption{ (Color online)  The blue points represent scatter plots for randomly prepared $X$ states evolved under under non-Markovian dephasing noise with $\mu=1$  for (a) $\eta_{p}=0.3 $ and (b) $\eta_{p}=1.0 $. The corresponding plots for $X$ states under the action of RTN in non-Markovian regime ($b=10\gamma$) for (c) $\gamma t =0.1$ (d) $\gamma t= 0.15$. (e) and (f) are the respective plots for OUN and PLN in the non-Markovian regime ($\Gamma= 16 \gamma$) with $\gamma t=0.15$. }
\label{fig3}
\end{figure}

As for the case of non-Markovian dephasing channels, here too we can see that the MNMSs retain their form but the Werner states and MEMSs lose their form and applicability. Thus, for the non-Markovian dephasing channel, the MNMSs always have the maximum amount of relative coherence $\mathcal{M}_0$ for any value of $\mathcal{M}_1$. Figure \ref{fig1} (b) shows the scatter plot of coherence of $X$ states for dephasing parameter ($\eta_p=0.3$) in the Markovian regime ($\alpha=0$) while  Figs. \ref{fig3} (a) and (b) illustrate the same for the dephasing parameters  ($\eta_p=0.3$ and $\eta_p=1$ respectively) in the non-Markovian regime ($\alpha=1$). We can clearly see from Fig. \ref{fig3} (b) that even in the case of $\eta_p=1$ the $X$ states have still some residual coherence left as per both measures of coherence while in the Markovian regime, the coherence would have disappeared. Therefore, we are able to observe the revival of coherence with an increase in dephasing parameter under the action of non-Markovian dephasing noise.  In fact, this is a characteristic feature of the non-Markovian channels where the coherence of the state first decreases with the progress of time, but again starts to increase due to back-flow of the information from the reservoir to the system. Figure \ref{fig2} (a) shows the rate of change of coherence as measured by $\mathcal{M}_0$ and $\mathcal{M}_1$ for an initial MNMS (Bell state).  It can be observed that $\mathcal{M}_1$ decays at a much faster rate in comparison to $\mathcal{M}_0$. Interestingly, we can observe that the rate of change of both $\mathcal{M}_0$ and $\mathcal{M}_1$ is not always  negative, as observed in the case of Markovian channels, but these rates can be positive as well. We can clearly see the revival of coherence with increase in dephasing parameter.

\subsubsection{Random telegraph noise}

Random telegraph noise (RTN) is another popular non-Markovian channel. This channel characterizes the dynamics when the system is subjected to a bi-fluctuating classical noise and the corresponding Kraus operators \cite{rtn1,naikoo1,naikoo2} are given by
\begin{eqnarray}\label{eq:RTN}
E_{0}^{\rm RTN} = \sqrt{\frac{1+\Lambda(t)}{2}} \sigma_{0},  \quad {\rm and} \quad
E_{1}^{\rm RTN} = \sqrt{\frac{1-\Lambda(t)}{2}} \sigma_{3}. 
\end{eqnarray}
Here, $\Lambda(t)$ is memory kernel and is given by
\begin{equation}\label{eq:lam}
\Lambda(t) = e^{-\gamma t} \left[ \cos \left( \gamma t \sqrt{\left(\frac{2b}{\gamma}\right)^2-1} \right) + \frac{\sin \left( \gamma t \sqrt{\left(\frac{2b}{\gamma}\right)^2-1} \right)}{\sqrt{\left(\frac{2b}{\gamma}\right)^2-1}}\right],
\end{equation}
where $b$ quantifies system-environment coupling strength, and $\gamma$ is proportional to the fluctuation rate of the RTN. The relative strength of the system-environment coupling and fluctuation rates decides the two regimes of RTN with a Markovian character if $4{b}^2 \leq \gamma^2$ and non-Markovian behavior otherwise.

The $X$ states evolved under RTN can be obtained using Eq.~(\ref{eq:RTN}) in Eq.~(\ref{Kr}), which attains the form as in Eq.~(\ref{x_state_PD}) with $\sqrt{\zeta}=\Lambda(t) $. Thus, $\mathcal{M}_0$ and $\mathcal{M}_1$ for the $X$ state under the action of RTN possess the same form as in Eqs.~(\ref{l1_state_PD}) and (\ref{crl_state_PD}) with $\sqrt{\zeta}=\Lambda(t)$. Notice that under non-Markovian regime, the function $\Lambda(t)$ in Eq.~(\ref{eq:lam}) has an oscillatory character due {to the presence of the} trigonometric functions {in the expression of $\Lambda(t)$,} in contrast to that in the Markovian regime where hyperboloc functions are found to be present. Therefore, we can clearly see that the coherence of the $X$ state decay monotonously in the Markovian regime, but in case of non-Markovian regime it shows revival of coherence. Hence, a periodic revival of coherence with time is observed, which is a peculiar feature of the non-Markovian channels.

Variation of coherence $\mathcal{M}_1$ with respect to $\mathcal{M}_0$ in this case is shown in Figs. \ref{fig3} (c) and (d) for $\gamma t=0.1$ and $0.15$, respectively.  We can clearly observe  that the parameters  $b=10\gamma$ correspond to the non-Markovian regime as $4{b}^2 > \gamma^2$ and thus coherence shows an oscillatory behavior with a periodic increase and decrease in its value with an increase in the damping factor. Also, the spread of the scatter plot can be observed to shrink with increasing noise, unlike Markovian channels, which can be {attributed to the} non-contraction behavior in this case.  The oscillating nature of coherence under the effect of the RTN channel in the non-Markovian regime is reflected in the rate of change of coherence as measured by measures $\mathcal{M}_0$ and  $\mathcal{M}_1$ for  an initial Bell state as shown in Fig. \ref{fig2} (a). The rate of change of relative entropy of coherence is faster than that of the $l_1$ norm of coherence, which is consistent with that observed so far under both Markovian and non-Markovian channels.

\subsubsection{Modified Ornstein-Uhlenback and power law noises}

The modified Ornstein-Uhlenback noise (OUN) and power law noise (PLN) have similar structure whose Kraus operators \cite{pln} are same as $E_{j}$s in Eq.~(\ref{eq:RTN}) with 
$\Lambda(t)$ given by 
\begin{equation}
\Lambda(t)= \begin{cases}
  \exp\left[ -\frac{\Gamma}{2} \left( t+ \frac{e^{\gamma t}-1}{\gamma}\right) \right], & \text{OUN } \\
  \exp\left[ \frac{(t \gamma +2)\Gamma t}{2(t \gamma +1)^2} \right], & \text{PLN}.
\end{cases}
\end{equation} 
Here, $\Gamma$ is the effective inverse relaxation time, and the spectral properties of the noise are given by $\gamma$ (bandwidth of noise). For large values of $\gamma$ the Markovian behavior is observed. In the present case, the $X$ states evolve to the same form as in Eq.~(\ref{x_state_PD}) with $\sqrt{\zeta}=\Lambda(t)$, obtained using Eq.~(\ref{eq:RTN}) in Eq.~(\ref{Kr}). Thus, $\mathcal{M}_0$ and $\mathcal{M}_1$ for the $X$ state under the effect of OUN and PLN channels are obtained as Eqs.~(\ref{l1_state_PD}) and (\ref{crl_state_PD}) with $\sqrt{\zeta}=\Lambda(t)$.

 Figures \ref{fig3} (e) and (f) show the scatter plot of the coherence as measured by the measures $\mathcal{M}_0$ and $\mathcal{M}_1$ for $X$ states under the action of OUN and  PLN, respectively. Unlike in the case of RTN, the presence of the non-Markovian regime here just leads to the slowing down of the decay of coherence. Particularly, the PLN is more effective in preserving the coherence of the initial state (cf. Fig. \ref{fig2} (a)). All of the remaining features are similar to that observed in the case of aforementioned Markovian and non-Markovian dephasing channels.

In this section, we have studied the evolution of $X$ states under some well-known dephasing-type Markovian and non-Markovian quantum channels. In such a channel, the eigenstates of the system do not change with time, but the phases evolve with time. Therefore, there is a loss of the information about the relative phases of the eigenstates with time. For such type of channels, only the off-diagonal elements of the evolved $X$ state possess the decoherence parameters, i.e., of the general form  given in Eq.~(\ref{x_state_PD}), and thus coherence measures $\mathcal{M}_0$ and $\mathcal{M}_1$ have a general form as given in Eqs.~(\ref{l1_state_PD}) and (\ref{crl_state_PD}) with $\zeta$.

This allows us to obtain a compact expression of the rate of change of both these measures with respect to the dephasing parameter $\zeta$ which revealed to us that the $l_1$ norm of coherence falls faster than the relative entropy of coherence for a state. This also leads to the conclusion that the dephased MNMSs retain their form and usefulness unlike the case of MEMSs and Werner states and thus have maximum $\mathcal{M}_0$ relative to $\mathcal{M}_1$ among the set of $X$ states.

In the next section, we study the evolution of $X$ states under another useful type of quantum channel, namely dissipative quantum channel.

\section{$X$ states under dissipative quantum channels} \label{sec3}

Quantum dissipative channels are the channels in which both the eigenstates as well as the relative phases between the eigenstates change with time. In this case, the energy of the system is lost to the reservoir and thus leading to a change in the population of system.

\subsection{$X$ states under quantum Markovian dissipative channels}

We begin with the evolution of $X$ states under some popular Markovian dissipative channels before discussing non-Markovian channels.

\subsubsection{Amplitude damping channel}
The amplitude damping channel is a schematic model of the decay of an excited state of a (two-level) atom due to spontaneous emission of a photon. The Kraus operators for a single qubit representing the amplitude damping channel \cite{nc} are 
\begin{equation}\label{AD-Kr}
E^{\rm AD}_0 = \left(
     \begin{array}{cc}
      1 & 0  \\
       0 & \sqrt{1-\eta_{a}}    
     \end{array}
   \right) \quad {\rm and} \quad
   E^{\rm AD}_1 = \left(
     \begin{array}{cc}
      0 & \sqrt{\eta_{a}} \\
       0 & 0     
     \end{array}
   \right),
\end{equation}
where $\eta_{a}$ is the damping parameter which is between $0$ and $1$ with $\eta_{a}=0$ representing noiseless channel. For two-qubit state $\rho$  in Eq.~(\ref{eq:Kr}) evolving under the amplitude damping channel, we can define the four Kraus operators  as $E^{\rm AD}_i \otimes E^{\rm AD}_j \, \forall i, j \in \{0,1 \}$.

To study the effect of amplitude damping channel acting on both the qubits of the $X$ states we use Eq.~(\ref{x_state}) as the initial state in Eq.~(\ref{eq:Kr}) with Eq.~(\ref{AD-Kr}) describing the corresponding Kraus operators. Therefore, the final state can be obtained as
\begin{equation} \label{x_state_AD}
\rho_{X}^{\rm AD} = \stackrel[\substack{i,j=0\\2i+j=k-1}]{1}{\sum} \rho_{kk}^{\rm AD}  \left|ij \right\rangle\left\langle ij \right| + \zeta \left( \rho_{14} \left|00 \right\rangle\left\langle 11 \right|+\rho_{23} \left|01 \right\rangle\left\langle 10 \right| +{\rm H.c.}\right)
\end{equation} 
with $\zeta=(1-\eta_{a})$ and
\begin{eqnarray}
\rho_{11}^{\rm AD} &= &\rho_{11} + (1-\zeta )(\rho_{22} + \rho_{33}+ (1-\zeta )\rho_{44}),\nonumber \\
\rho_{44}^{\rm AD} &= & \zeta ^{2}\rho_{44},\nonumber \\
\rho_{jj}^{\rm AD} &= & \zeta (\rho_{jj}+(1-\zeta )\rho_{44}) \, \forall\, j\in \{2,3 \}.\label{rho-AD}
\end{eqnarray}
Notice that in contrast to the present case, the diagonal elements were $\zeta$ independent in Eq.~(\ref{x_state_PD}) which represents the form obtained for the $X$ state evolving over phase damping channel and the other dephasing channels.

\begin{figure}[tb]
\includegraphics[width=\textwidth]{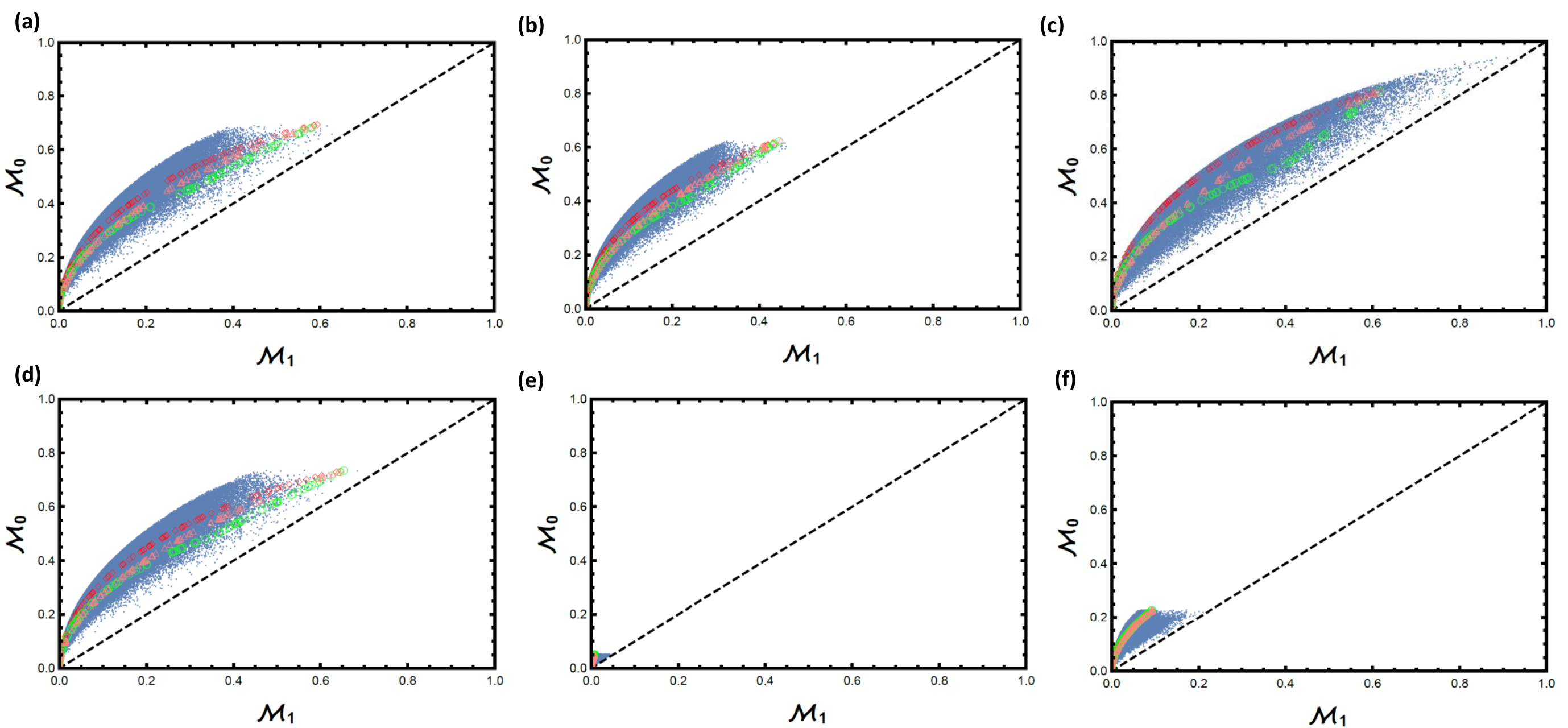}
\caption{ (Color online) The blue points represent scatter plots for $X$ states under (a) amplitude damping with $\eta_{a}=0.3 $, (b) combined phase and amplitude damping noise with $\eta_{a}=0.3,\,\eta_{p}=0.1$, (c) amplitude damping with memory with $\eta_{a}=0.3$ and $\mu=0.9 $. $X$ states under non-Markovian amplitude damping channel in the non-Markovian regime ($\Gamma=0.1\gamma$) for (d) $\gamma t=0.25 $, (e) $\gamma t=1$ and (f) $\gamma t=1.5 $.}
\label{fig4}
\end{figure}

Interestingly, $\mathcal{M}_0$ for the resultant state $\rho_{X}^{\rm AD}$ under the presence of amplitude damping channel resulted in the similar expression as Eq.~(\ref{l1_state_PD}) with $\zeta=(1-\eta_{a})$. 
Similarly, $\mathcal{M}_1$ for the resultant state $\rho_{X}^{\rm AD}$ evolved over the amplitude damping channel is computed as
\begin{equation}\label{crl_state_AD}
\mathcal{M}_1(\rho _{X}^{\rm AD})= \sum_{i} \lambda _{i}^{\rm AD} \log_{2}(\lambda_{i}^{\rm AD})- \sum_{i} \rho _{ii}^{\rm AD} \log_{2}(\rho _{ii}^{\rm AD})  , 
\end{equation}
where $\lambda _ {i}^{\rm AD} $s are the eigenvalues of the state $\rho_{X}^{\rm AD}$ as given in Eq.~(\ref{EV}). As the diagonal elements of $\rho_{X}^{\rm AD}$ depend upon $\zeta$ so the second term in the expression of $\mathcal{M}_1$ in Eq.~(\ref{crl_state_AD}) differs from that in Eq.~(\ref{crl_state_PD}). In fact, this is the case for all dissipative type noise which makes it difficult to get a compact expression for $\mathcal{M}_1$ in terms of the initial elements of the $X$ state.

Clearly we can see that the effect of amplitude damping on the coherence of $X$ states is to reduce $\mathcal{M}_0$ monotonically by a factor of $(1-\eta_{a})$. This is similar to that for the case of phase damping where the  $\mathcal{M}_0$ reduces by a factor of $(1-\eta_{p})$. So, it seems that the effect of amplitude damping channel on $\mathcal{M}_0$ is similar to that for the case of all dephasing type noises where the $\mathcal{M}_0$ reduces by a factor of damping parameter $\zeta$ with respect to the initial state. However, the effect of dissipative type noise is different in case of coherence measured by $\mathcal{M}_1$. 

Figure \ref{fig4} (a) represents the scatter plots of the coherence of $X$ states as calculated via $\mathcal{M}_0$ on $y$ axis and $\mathcal{M}_1$ on the $x$ axis for the amplitude  damping coefficient $\eta_{a}=0.3$. The scatter plot of coherence for the initial $X$ states as measured by $\mathcal{M}_0$ and $\mathcal{M}_1$ in the absence of any noise, which {corresponds to amplitude damping case $\eta_{a}=0$,} is shown in Fig. \ref{fig1} (a). In contrast to the dephasing type noises, we can see that under the action of amplitude damping channel, the deformed MNMSs (\ref{mnms_state}) no longer retain their form and hence no longer form the boundary states with maximum value of $\mathcal{M}_0$ for a given value of $\mathcal{M}_1$. Further, we can conclude from Fig. \ref{fig4} (a) that as the amount of damping coefficient $\eta_{a}$ starts increasing, the MNMSs (\ref{mnms_state}), MEMSs (\ref{mems_state}) and Werner states  (\ref{werner_state}) with amplitude damping noise starts to come closer due to contraction property of the CPTP map \cite{NM-rev}.  We can further observe that with an increase in damping parameter $\eta_a$ the maximum attainable coherence reduces with respect to both the measures of coherence. Consequently, the length of the region occupied by the scattered plot reduces. Figure \ref{fig2} (b) represents the rate of change of coherence as measured by the two measures $\mathcal{M}_0$ and $\mathcal{M}_1$ for an initial Bell state. As expected, we can see that rate of change of $\mathcal{M}_1$ is faster than the $\mathcal{M}_0$. Interestingly, $\mathcal{M}_0$ falls at the same rate for both amplitude and phase damping while $\mathcal{M}_1$ falls more rapidly under the effect of phase damping channel (cf. Fig. \ref{fig2}). 

We further consider the effect of some other Markovian and non-Markovian dissipative type noises.

\subsubsection{Combined amplitude and phase damping channel}

The Kraus operators for the combined effect of amplitude and phase damping \cite{baan} acting on a single qubit is given by 
\begin{eqnarray}\label{AP-Kr}
E_0 ^{\rm AP}=& \left(
     \begin{array}{cc}
      1 & 0  \\
       0 & \sqrt{(1-\eta_{p})(1-\eta_{a})}    
     \end{array}
   \right), &
   E_1^{\rm AP} = \left(
     \begin{array}{cc}
      1 & 0 \\
       0 & \sqrt{\eta_{p} (1-\eta_{a})}     
     \end{array}
   \right),\nonumber \\
E_2 ^{\rm AP}=& \left(
     \begin{array}{cc}
      1 & \sqrt{\eta_{a}(1-\eta_{p})}  \\
       0 & 0   
     \end{array}
   \right), &
   E_3^{\rm AP} = \left(
     \begin{array}{cc}
      1 & 0 \\
       0 & \sqrt{\eta_{p} \eta_{a}}     
     \end{array}
   \right).
\end{eqnarray}
If we consider the same noise to be acting on both the qubits, then the Kraus operators is given by the tensor product of the individual Kraus operators as  $E^{\rm AP}_i\otimes E^{\rm AP}_j$ with $i,j= 0,1,2,3$.

The resultant $X$ state (\ref{x_state}) under the combined effect of amplitude and phase damping channel can be computed using Eq.~(\ref{AP-Kr}) in Eq.~(\ref{Kr}) as:
\begin{equation} \label{x_state_APD}
\rho_{X}^{\rm AP} = \stackrel[\substack{i,j=0\\2i+j=k-1}]{1}{\sum} \rho_{kk}^{\rm AD}  \left|ij \right\rangle\left\langle ij \right| + \zeta \left( \rho_{14} \left|00 \right\rangle\left\langle 11 \right|+\rho_{23} \left|01 \right\rangle\left\langle 10 \right| +{\rm H.c.}\right)
\end{equation} 
with $\zeta=(1-\eta_{a})(1-\eta_{p})$ and $\rho_{kk}^{\rm AD}$ defined in Eq.~(\ref{rho-AD}). The presence of this channel shows the characteristic features of both amplitude damping as well as that of phase damping. Measure $\mathcal{M}_0$ for the resultant state $\rho_{X}^{\rm AP}$ states under the presence of phase damping channel can be computed as $\mathcal{M}_0(\rho_{X}^{\rm AP})= \zeta \mathcal{M}_0(\rho_{X})$. Therefore, $\mathcal{M}_0$ for the $X$ states under the combined effect of amplitude and phase damping channel reduces the coherence by a factor of $\zeta$. 

Figure \ref{fig4} (c) represents the scatter plot of the coherence of $X$ states with $\mathcal{M}_0$ on the $y$ axis and $\mathcal{M}_1$ on the $x$ axis for a particular value of $\zeta$. Similar to the previous observations, we can see that the area under the scatter plot decreases due to the decrease in the coherence as measured by $\mathcal{M}_0$ and $\mathcal{M}_1$. Due to the combined effect of phase and amplitude damping the rate of decrease of coherence is observed to be faster compared to their independent effect. Further, we can see that the evolved MNMSs under the presence of combined action of phase and amplitude damping noise no more form the boundary states for the relative coherence.

Hereafter, we consider the effect of addition of memory component in the amplitude damping noise.

\subsubsection{Amplitude damping with memory}

For amplitude damping with memory \cite{AD-mem}, the uncorrelated noise in the two-qubit state (\ref{eq:mem}) is given by
\begin{equation}
E^{u}_{i,j}= E^{\rm AD}_i \otimes E^{\rm AD}_j\,\forall (i,j = 0,1),
\end{equation}
where $E^{\rm AD}_k$ is defined in Eq.~(\ref{AD-Kr}).

The Kraus operators for the correlated noise \cite{AD-mem} is given by 
\begin{equation}
E^{c}_{00} =\left|00 \right\rangle\left\langle 00 \right|+\sqrt{1-\eta_{a}} \left|11 \right\rangle\left\langle 11 \right| \quad {\rm and} \quad
  E^{c}_{11} =\sqrt{\eta_{a}}\left|00 \right\rangle\left\langle 11 \right|
\end{equation}
with $\eta_{a}$ as amplitude damping parameter.

The resultant $X$ state under the action of amplitude damping channel with memory acting on both the qubits can be written as
\begin{equation} \label{x_state_ADM}
\rho_{X}^{\rm ADM} = \stackrel[\substack{i,j=0\\2i+j=k-1}]{1}{\sum} \rho_{kk}^{\rm ADM}  \left|ij \right\rangle\left\langle ij \right| +  \left( \rho_{14}^{\rm ADM} \left|00 \right\rangle\left\langle 11 \right|+\rho_{23}^{\rm ADM} \left|01 \right\rangle\left\langle 10 \right| +{\rm H.c.}\right),
\end{equation} 
where the elements are
\begin{eqnarray}\label{eq:ADMte}
\rho_{11}^{\rm ADM} &=& \rho_{11}^{\rm AD} -\mu\left(1-\zeta\right)  \left( \rho _{22}+\rho _{33}-\zeta\rho _{44}\right), \\ \nonumber 
\rho_{44}^{\rm ADM} &=& \rho_{44}^{\rm AD}+ \mu \zeta \left(1-\zeta\right) \rho _{44}, \\ \nonumber
\rho_{jj}^{\rm ADM} &=& { \rho_{jj}^{\rm AD}+\mu\left(1-\zeta\right)  \left(\rho _{jj}- \zeta \rho _{44}\right)}  \, \forall\, j\in \{2,3 \},\\ \nonumber
\rho_{14}^{\rm ADM} &=& \left(\zeta+\mu \sqrt{\zeta}(1-\sqrt{\zeta}) \right)  \rho _{14}, \\ \nonumber
\rho_{23}^{\rm ADM} &=& \left(\zeta+\mu \eta _a\right) \rho _{23}
\end{eqnarray}
with $\zeta=(1-\eta_{a})$ and $\rho_{kk}^{\rm AD}$ defined in Eq.~(\ref{rho-AD}). Thus, $\mathcal{M}_0$ can be obtained as
\begin{equation} \label{l1_state_ADM}
\mathcal{M}_0(\rho_{X}^{\rm ADM})= 2 \{ \vert \rho_{14}^{\rm ADM} \vert + \vert  \rho_{23}^{\rm ADM} \vert \} = \left(\zeta+\mu\eta _a\right) \mathcal{M}_0(\rho_{X})+2\mu (\sqrt{\zeta}-1)\vert \rho _{14}\vert.
\end{equation}

If the memory component of the channel is zero (i.e., $\mu=0$), the state $\rho_{X}^{\rm ADM}$ would reduce to that of  amplitude damping channel (\ref{x_state_AD}) and we can reduce all the results obtained in that section. We can clearly observe from Fig. \ref{fig4} (b) that the features seen from application of amplitude damping channel are also observed for $\mu \neq 0$. The presence of memory component ($\mu \neq 0$) has the effect of slowing down the diminishing of coherence. With an increase in the value of memory component ($0 \le \mu \le 1 $), the coherence of the system once again starts increasing. This is due to the fact that memory component is related to the existence of correlation between two consecutive uses of the channel. It is interesting to observe a stark difference between amplitude damping channel with memory in comparison to the earlier mentioned phase damping with memory. For the case of phase damping in the presence of full memory ($\mu=1$), we observed that the $X$ states did not decohere at all for any value of damping factor ($\eta_p$) but an analogous feature is not observed for the case of amplitude damping with full memory ($\mu=1$).  Notice in Eq.~(\ref{eq:ADMte}) that $\rho_{23}^{\rm ADM}=\rho_{23}$ while $\rho_{14}^{\rm ADM}=0$ for $\mu=1=\eta_a$. This is due to the fact that over amplitude damping noise, there is always a loss of energy from the system to the environment and this loss of energy can never be perfectly stalled by an application of correlations between the consecutive uses of the channel.

In what follows, we discuss the effect of a non-Markovian dissipative noise.

\subsection{$X$ states under quantum non-Markovian dissipative channels}

Finally, we discuss the effects of non-Markovian dissipative channels on coherence of the $X$ states.

\subsubsection{Non-Markovian amplitude damping}

The Kraus operators for a single qubit under the effect of the non-Markovian dissipative noise \cite{nmad1,nmad2} are 
\begin{equation}\label{eq:NMAD}
E^{\rm NMAD}_0 = \left(
     \begin{array}{cc}
      1 & 0  \\
       0 & \sqrt{p(t)}    
     \end{array}
   \right) \quad {\rm and} \quad
   E^{\rm NMAD}_1 = \left(
     \begin{array}{cc}
      0 & \sqrt{1-p(t)} \\
       0 & 0     
     \end{array}
   \right).
\end{equation}
Here, $p(t)$ is known as memory kernel and is given by 
\begin{equation}
p(t) = e^{-\Gamma t} \left[ \cos \left( \sqrt{(2 \gamma \Gamma -\Gamma^2)} \frac{t}{2} \right) + \frac{\Gamma}{\sqrt{(2 \gamma \Gamma -\Gamma^2)}} \sin \left( \sqrt{(2 \gamma \Gamma -\Gamma^2)} \frac{t}{2} \right)\right]^{2},
\end{equation}
where $\gamma$ quantifies system-environmental coupling strength and is related to qubit relaxation time while $\Gamma$ is the line width which depends on the reservoir correlation time. The non-Markovian effects start playing their role when the reservoir correlation time becomes greater than the qubit relaxation  time. { Thus, the value of parameter $({2 \gamma }/{\Gamma})$ decides the two regimes of this noise with a Markovian character if ${2 \gamma}\leq{\Gamma}$ while non-Markoian otherwise.} The evolved state obtained using Eqs.~(\ref{Kr}) and (\ref{eq:NMAD}) possess the same form as given by Eq.~(\ref{x_state_AD}) with $\zeta=p(t)$. 

Measure $\mathcal{M}_0$ for the $X$ state under the action of non-Markovian amplitude damping noise can be computed as $\mathcal{M}_0(\rho_{X}^{\rm NMAD})= \zeta \mathcal{M}_0(\rho_{X})$. So, the amount of coherence is dependent upon the behavior of the memory kernel $p(t)$ which has two different regimes. In the non-Markovian regime, we can observe a periodic decay and revival of the coherence.  Figures \ref{fig4} (d), (e) and (f) represent the scatter plots for the coherence measured by two measures $\mathcal{M}_0$ and $\mathcal{M}_1$ for different values of decay parameter $\gamma t$. We can conclude that the coherence first decrease and then starts increasing again with an increase in $\gamma t$. Further, the oscillating behavior of the memory kernel is reflected in the rate of change of coherence as measured by $\mathcal{M}_0$ and $\mathcal{M}_1$ for an initial Bell state as shown in Fig. \ref{fig2} (b).

In this section, we have studied evolution of $X$ states under dissipative type Markovian and non-Markovian quantum channels. For such type of channels, both diagonal and off-diagonal elements of the evolved $X$ states possess the decoherence parameters, i.e., for instance given in Eq.~(\ref{x_state_AD}), and thus coherence measures $\mathcal{M}_0$ and $\mathcal{M}_1$ have a general form as Eqs.~(\ref{l1_state_PD}) and (\ref{crl_state_AD}) with $\zeta$.
In the next section, we will discuss about the usefulness of the $X$ states in the presence of noise for the purpose of quantum teleportation.

\section{Quantum teleportation fidelity for $X$ states used as a resource } \label{sec4}

Teleportation \cite{tp1,tp2} is the process of transferring the state of an unknown qubit from one party (Alice) to another party (Bob) via the use of a shared entangled resource. The performance of teleportation is quantified in terms of fidelity, which represents how close the state received by Bob is with respect to that sent by Alice. With the use of a pure maximally entangled state (Bell state) shared between Alice and Bob, we can devise a way by which we can teleport the unknown state with unit fidelity. However, the use of mixed state  invariably leads to a reduction in the fidelity of the teleported state. Further, it has been shown that one can teleport a quantum state by use of only classical resources, but the maximum achievable fidelity is $2/3$ (classical limit) \cite{pop}. Therefore, an arbitrary quantum state is useful for quantum teleportation task only if it can teleport an unknown state with fidelity greater than the classical limit. 
{In what follows, we examine the relation of coherence of $X$ states with the performance of teleportation under the detrimental effect of noise.}

\begin{figure}[tb]
\includegraphics[width=\textwidth]{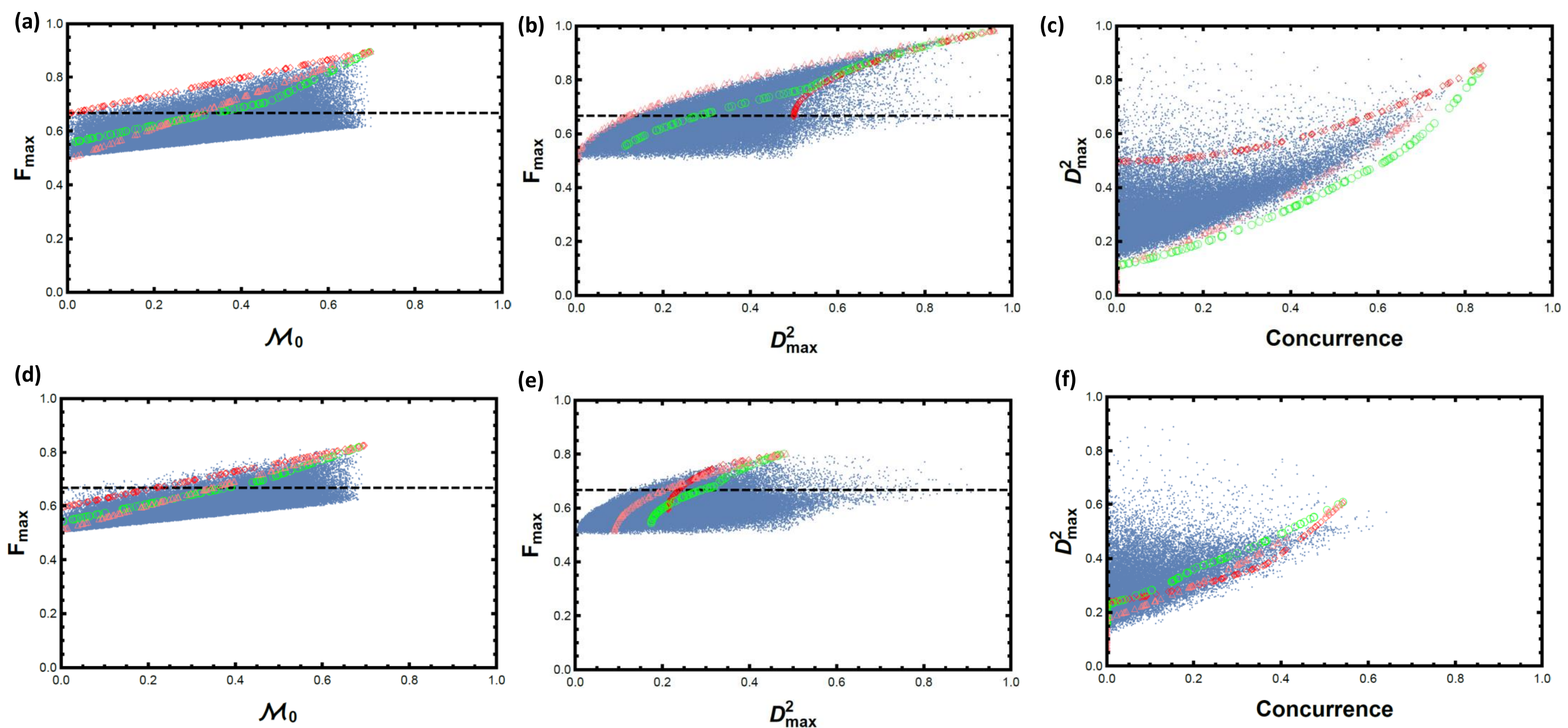}
\caption{ (Color online) { The blue points represent scatter plots for $X$ states under (a)-(c) dephasing and (d)-(f) dissipative type channels. We show in (a) phase damping with $\eta_p=0.3$, (b) phase damping with memory at $\eta_p=0.3,\, \mu=0.9$,  (c) RTN with $b=10\gamma$ and $\gamma t= 0.15$, (d) amplitude damping with $\eta_a=0.3$, (e) combined amplitude and phase damping with $\eta_a=0.3,\,\eta_p=0.1 $, (f) non-Markovian amplitude damping noise with $\Gamma= 0.1\gamma$ and $\gamma t=0.25 $. The black (dashed) line represents the classical limit of fidelity $2/3$. }}
\label{fig5}
\end{figure}

In the process of teleportation, it is known that if Bob is equipped to perform all kinds of unitary transformations to the qubit at his possession, then the maximal average fidelity achievable can be expressed as \cite{f1max,hu_tele, werner2}:
\begin{equation}
F_{\max}= \frac{1}{2}+ \frac{1}{6} N(\rho),
\end{equation}
where $N(\rho)$ for the $X$ states \cite{hu_tele} can be obtained as $N(\rho)= 2 [| \rho_{14}| + |\rho_{23}| + |(|\rho_{14}| - |\rho_{23}|)|] + |\rho_{11} + \rho_{44} - \rho_{22} -\rho_{33}|$. We can clearly see that for all the states with $N(\rho)\ge 1$, the maximum fidelity $F_{\max}\ge 2/3$, i.e., such states are useful for quantum teleportation. Further, it is known that all the bipartite states which violate the CHSH Bell inequality are useful for quantum teleportation \cite{f1max}. For $X$ states, Bell nonlocality requires
\begin{equation}
M(\rho) = \max \{ 8 (|\rho_{14}|^{2} + |\rho_{23}|^{2}), 4 (|\rho_{14}| + |\rho_{23}|)^{2}+ (\rho_{11} + \rho_{44} - \rho_{22} -\rho_{33})^{2} \}>1,
\end{equation}
where $M(\rho)$ corresponds to the degree of violation of the Bell inequality for a bipartite state. Among the set of all quantum states in the form of $X$ states, MNMSs are always useful for quantum teleportation as they always violate CHSH inequality. {Due to the same reason,  MNMSs would also be useful in realizing device independent quantum key distribution in noisy environment.}

 We can further obtain the expression for $F_{\max}$ for the case of MNMSs under dephasing type channels. Maximal average fidelity for MNMSs (\ref{mnms_state}) under dephasing type channels can be obtained in terms of $N(\rho)$ as
\begin{equation}
N(\rho_{X}^{\rm MNMS})= 1 + \mathcal{M}_0(\rho^{\rm MNMS}_{X}).
\end{equation}
Since $\mathcal{M}_0(\rho^{\rm MNMS}_{X}) \ge 0$ we always have $N(\rho_{X}^{\rm MNMS}) \ge 1$, and hence maximal average fidelity $F_{\max}$ is greater than the classical limit (cf. Fig. \ref{fig5} (a)). In other words, MNMSs always remain useful for quantum teleportation under dephasing type channels because a MNMS transforms to another MNMS under the set of dephasing channels studied here.

Similarly, for MNMSs evolving under dissipative type channels gives $N(\rho)$ as
\begin{equation}
N(\rho_{X}^{\rm MNMS})=  \mathcal{M}_0(\rho^{\rm MNMS}_{X}) + (\rho^{\prime}_{11}+\rho^{\prime}_{44}-\rho^{\prime}_{22}-\rho^{\prime}_{33}),
\end{equation} 
where $\rho^{\prime}_{ii} $ are the population elements of the MNMSs after decoherence. Thus, the value $N(\rho_{X}^{\rm MNMS})$ depends both on the initial population and the dissipative noise parameter(s). Therefore, unlike dephasing channels, for certain values of the parameters the MNMS channels may have maximal average teleportation fidelity less than $2/3$ (see Fig. \ref{fig5} (d)). For the rest of the classes of $X$ states, their usefulness in quantum telportation depends on both the initial state and noise parameters for both dephasing and dissipative type noises. 

Independent of the resource theoretic coherence measures, Jiri et al.~introduced first-order coherence \cite{optical_coh} which is based on the intrinsic distribution of coherence between the sub-systems. In two-qubit scenario, the coherence for each subsytem is defined as $D_{i} = \sqrt{2\,Tr[\rho_{i}^2]-1}\,\, \forall\, i\in \left\{A,B\right\} $, and the amount of coherence for the complete system is $ D^{2}= (D_{A}^{2}+D_{B}^{2})/2$. 
The amount of first-order coherence can be maximized for a quantum state $\rho _{AB}$ with the help of a global unitary ($U$) operation such that $\rho^{\prime}_{AB}= U\rho_{AB}U^{\dagger}$. This gives $ D^{2}_{\max} = (\lambda _{1}- \lambda _{4})^{2}+(\lambda _{2}- \lambda _{3})^{2}$ known as hidden coherence which corresponds to the maximum attainable coherence due to global unitary transformations,  where $\lambda_i$s are the eigenvalues of the state $\rho _{AB}$ in a decreasing order \cite{optical_coh}. Figure \ref{fig5} (b) shows the scatter plot of the $F_{\max}$ with  $ D^{2}_{\max}$  for dephasing type noise (phase damping with memory)  while Fig. \ref{fig5} (e) shows the same for a dissipative type noise (amplitude damping channel with memory). The minimum value of $ D^{2}_{\max}$ in the former case for the MNMS, MEMS and Werner state are $0.5$, $0.1$ and $0$, respectively. Further, we can clearly see that for a given value of $ D^{2}_{\max}$ of the $X$ states after decoherence, the evolved Werner states always have the maximum amount of $F_{\max}$. However, this feature can not be observed in any of the dissipative type noises as we can observe the $X$ states with the same amount of $ D^{2}_{\max}$ can have $F_{\max}$ more/less than that of the Werner state. In our previous work \cite{mishra_x}, we have established a relation between the $ D^{2}_{\max}$ and the amount of entanglement as measured by concurrence for the $X$ states and observed that the MEMSs have the lowest amount of concurrence for a given amount of $ D^{2}_{\max}$. Under the action of different dephasing and dissipative type noises, we can observe from Figs. \ref{fig5} (c) and (f)  that the decohered MEMSs still form the lower boundaries for the scatter plot between $ D^{2}_{\max}$ and concurrence for the dephasing type noise, but not for the dissipative type noises. The study of relations between $ D^{2}_{\max}$ with the teleportation fidelity $F_{\max}$ and the concurrence has provided a completeness to our study. 

\section{Conclusion} \label{sec5}

We studied the evolution of coherence for $X$ states under the presence of different type of noises. We have observed the characteristic traits of the differently studied Markovian and non-Markovian channels. Interestingly, the action of all the studied popular types of noises, $X$ states always retain their form, hence all the results obtained in our previous study \cite{mishra_x} remain equally valid here, too. Further, we have divided the noise into two broad categories namely dephasing type noises (where there is no loss of energy to the environment) and dissipative type noise (where there is a loss of energy from the system to the environment). For dephasing type noise, we observed that the MNMSs even after decoherence always form the upper boundary of the scatter plot of the relative coherence measured by $l_1$ norm and relative entropy of coherence due to the fact that under dephasing type noise an arbitrary MNMS always maps to a MNMS only. Hence, all MNMSs will be useful to teleport a qubit with maximal average fidelity greater than the classical limit $2/3$. Further, MNMS will also be useful for device independent quantum key distribution. However, this feature is not to be observed in any of the dissipative type noises as there is a change is the population elements, too. For the rest of the states under both dephasing and dissipative type noises, the usefulness in quantum teleportation is dependent on the initial state of the system as well as the noise parameters. In general, the coherence measured through relative entropy of coherence decays faster than that using $l_1$ norm of coherence. Further, we observed that the presence of the full memory component $\mu =1$ in phase damping channel with memory results in stalling the decoherence. Therefore, we have a situation where there is a  freezing of the coherence and all other parameters of the system.  This situation is not observed in any of the other type of noises though we were able to see the slowing down or the revival of coherence under the effect of non-Markovian noise. Apart from the resource theoretic measures of coherence, we also studied that interconnections between the hidden coherence $(D^{2}_{\max})$ with the maximal achievable teleportation fidelity $(F_{\max})$. We found that the decohered Werner state forms the upper boundary of the scatter plot of $F_{\max}$ with $D^{2}_{\max}$ for all dephasing type noises, but not so for the case of any of the dissipative type noise. Further, the MEMSs after decoherence under dephasing type noises form the lower boundary of the scatter plot of concurrence with $D^{2}_{\max}$ as was the case for our study of $X$ states in ideal case \cite{mishra_x} but not so in any of the dissipative type noises. Our study of the coherence of the $X$ states under the presence of various types of quantum channels provides a clue towards their applications in the realistic physical scenarios. We hope the present study will help in developing methods to maintain and exploit the the quantum coherence in scalable quantum computers and possible applications.

\textbf{Acknowledgment}
AP and SM acknowledge the support from the QUEST scheme of Interdisciplinary
Cyber Physical Systems (ICPS) program of the Department of Science and Technology (DST), India (Grant
No.: DST/ICPS/QuST/Theme-1/2019/14). KT acknowledges GA \v{C}R (project No.
18-22102S) and support from ERDF/ESF project `Nanotechnologies for Future'
(CZ.02.1.01/0.0/0.0/16\_019/0000754).


\begin{thebibliography}{99}


\bibitem{Mandel} L. Mandel and E. Wolf, \textit{Optical Coherence and Quantum
Optics} (Cambridge University Press, Cambridge, 1995).

\bibitem{glaub} R. J. Glauber, Phys. Rev. \textbf{131}, 2766 (1963).

\bibitem{sudar}E. C. G. Sudarshan, Phys. Rev. Lett. \textbf{10}, 277 (1963).

\bibitem{coh_rev2} A. Streltsov, G. Adesso, and M. B. Plenio, Rev. Mod. Phys. \textbf{89}, 041003 (2017).

\bibitem{coh_rev1} M.-L. Hu, X. Hu, Y. Peng, Y.-R. Zhang, and H. Fan, Phys. Rep. \textbf{762}, 1 (2018).

\bibitem{bio1} N. Lambert, Y. N. Chen, Y. C. Cheng, C. M. Li, G. Y. Chen, and F. Nori, Nature Physics. \textbf{9}, 10 (2013).

\bibitem{bio2}  E. M. Gauger,  E. Rieper, J. J. L. Morton, S. C. Benjamin, and V. Vedral, Phys. Rev. Lett. \textbf{106}, 040503 (2011).

\bibitem{bio3} S. Lloyd, J. Phys. Conf. Ser. \textbf{302}, 012037 (2011).


\bibitem{plenio} T. Baumgratz, M. Cramer, and M. B. Plenio, Phys. Rev. Lett. \textbf{113}, 140401 (2014).


\bibitem{winter} A. Winter and D. Yang, Phys. Rev. Lett. \textbf{116}, 120404 (2016).

\bibitem{bera} M. N. Bera, T. Qureshi, M. A. Siddiqui, and A. K. Pati, Phys. Rev. A \textbf{92}, 012118 (2015).

\bibitem{sandeep1} A. Venugopalan, S. Mishra, and T. Qureshi, Physica A \textbf{516}, 308 (2019).

\bibitem{sandeep2} S. Mishra, A. Venugopalan, and T. Qureshi,  Phys. Rev. A \textbf{100}, 042122 (2019).

\bibitem{mishra_x} S. Mishra, K. Thapliyal, A. Pathak, and  A. Venugopalan, Quantum Inf. Process.  \textbf{18}, 295 (2019).

\bibitem{optical_coh} J. Svozil\'{\i}k, A. Vall\'es, J. Pe\v{r}ina Jr, and J. P. Torres, Phys. Rev. Lett. \textbf{115}, 220501 (2015).

\bibitem{coh-res} K. C. Tan, T. Volkoff, H. Kwon, and H. Jeong, Phys. Rev. Lett. \textbf{119}, 190405 (2017).

\bibitem{xstate1} T. Yu and J. H. Eberly, Phys. Rev. Lett. \textbf{93}, 140404 (2004).

\bibitem{xstate2} T. Yu and J. H. Eberly, Quantum Inform. Comput. \textbf{7}, 459 (2007).

\bibitem{esd} T. Yu and J. H. Eberly, Science \textbf{323}, 598 (2009).

\bibitem{x_dynamics} N. Quesada , A. Al-Qasimi, and D. F. V. James, J. Mod. Opt. \textbf{59}, 1322 (2012).

\bibitem{werner} R. F. Werner, Phys. Rev. A \textbf{40}, 4277 (1989).

\bibitem{mnms} J. Batle and M. Casas, J. Phys. A : Math. Theor. \textbf{44}, 445304 (2011).

\bibitem{ishizaka} S. Ishizaka and T. Hiroshima, Phys. Rev. A \textbf{62}, 022310 (2000).

\bibitem{verst} F. Verstraete, K. Audenaert, and B. D. Moor, Phys. Rev. A \textbf{64}, 012316 (2001).

\bibitem{munro} W. J. Munro, D. F. V. James, A. G. White, and P. G. Kwiat, Phys. Rev. A  \textbf{64}, 030302 (2001).

\bibitem{xh}  P. E. Mendonça, M. A. Marchiolli, and D. Galetti, Ann. Phys. \textbf{351}, 79 (2014).

\bibitem{xp_p1} N. A. Peters, J. B. Altepeter, D. A. Branning, E. R. Jeffrey, T. C. Wei, and P. G. Kwiat, Phys. Rev. Lett. \textbf{92}, 133601 (2004).

\bibitem{xp_p2} M. Barbieri, F. De Martini, G. Di Nepi, and P. Mataloni, Phys. Rev. Lett. \textbf{92}, 177901 (2004).

\bibitem{xp_p3} J. B. Altepeter, D. Branning, E. Jeffrey, T. C. Wei, P. G. Kwiat, R. T. Thew , J. L. O’Brien, M. A. Nielsen, and A. G. White, Phys. Rev. Lett. \textbf{90}, 193601 (2003).

\bibitem{xp_p4} Y. S. Zhang, Y. F. Huang, C. F. Li, and  G. C. Guo, Phys. Rev. A \textbf{66}, 062315 (2002).

\bibitem{xp_uc1}  C. Monroe, D. M. Meekhof, B. E. King,  W. M. Itano, and D. J. Wineland, Phys. Rev. Lett. \textbf{75}, 4714 (1995).

\bibitem{xp_uc2} G. S. Agarwal and K.T. Kapale, Phys. Rev. A \textbf{73}, 022315 (2006).

\bibitem{xp_nmr} A. R. P. Rau, Phys. Rev. A \textbf{61}, 032301 (2000).

\bibitem{x2021} W.F. Balthazar, D. G. Braga, V. S. Lamego, M. M. Passos, and J. A.  Huguenin, Phys. Rev. A \textbf{103}, 022411 (2021).

\bibitem{nc} M. A. Nielsen and I. L. Chuang, \textit{Quantum Computation and Quantum Information} (Cambridge University Press, Cambridge 2000).

\bibitem{pet_book} H. P. Breuer and F. Petruccione, \textit{The Theory of Open Quantum Systems} (Oxford University Press, New York, 2002).

\bibitem{QC-rev}F. Caruso, V. Giovannetti, C. Lupo, and S. Mancini, Rev. Mod. Phys. \textbf{86}, 1203 (2014).

\bibitem{kraus} K. Kraus, Ann. Phys. (N. Y.) \textbf{64}, 311 (1971).

\bibitem{NM-rev} H. P. Breuer, E. M. Laine, J. Piilo, and B. Vacchini, Rev. Mod. Phys. \textbf{88}, 021002 (2016).

\bibitem{NM-mem} M. Ban, Phys. Rev. A  \textbf{99}, 012116 (2019).

\bibitem{radha2019} C. Radhakrishnan, Z. Lü, J. Jing, and T. Byrnes, Phys. Rev. A \textbf{100}, 042333 (2019).

\bibitem{young2020}  J. D. Young and  A. Auyuanet,  Quantum Inf. Process. \textbf{19}, 398 (2020).

\bibitem{song2020} Y. Song, Y. Wang, H. Tang, and Z. Zhao, Int. J. Theor. Phys. \textbf{59}, 873 (2020)

\bibitem{zhao2020} M. J. Zhao, T. Ma, Z. Wang, S. M. Fei, and R. Pereira, Quantum Inf. Process. \textbf{19}, 104 (2020). 

\bibitem{wang2019} Y. S. Wang, D. Wang, and L. Ye, Int. J. Theor. Phys. \textbf{58}, 3667 (2019).

\bibitem{frozen2015} T. R. Bromley, M.  Cianciaruso, and G. Adesso,  Phys. Rev. Lett. \textbf{114}, 210401 (2015).

\bibitem{huang2017} Z. Huang and H. Situ, Quantum Inf. Process. \textbf{16}, 222 (2017).

\bibitem{guo} Y. N. Guo, Q. L. Tian and K. Zeng, Quantum Inf. Process. \textbf{16}, 310 (2017).

\bibitem{liu2017} C. L. Liu, Y. Q.  Guo and D. M. Tong, Phys. Rev. A \textbf{96}, 062325 (2017).

\bibitem{luo2019} S. Luo and Y. Sun, Phys. Lett. A \textbf{383}, 2869 (2019).

\bibitem{jiang2020} Z. Jiang, T.  Zhang, X.  Huang and S. M. Fei, Quantum Inf. Process. \textbf{19}, 92 (2020).

\bibitem{zhang2019} C. Zhang, T. R. Bromley, Y. F. Huang, H. Cao, W. M. Lv, B. H.  Liu, C. F. Li, G. C. Guo, M. Cianciaruso, and G. Adesso, Phys. Rev. Lett.  \textbf{123}, 180504 (2019).

\bibitem{cai2020} X. Cai, Sci. Rep. \textbf{10}, 88 (2020).

\bibitem{naikoo1} J. Naikoo and S.  Banerjee,  Quantum Inf. Process.  \textbf{19}, 29 (2020).

\bibitem{naikoo2} J. Naikoo, S. Banerjee,  and C. M. Chandrashekar,  Phys. Rev. A \textbf{102}, 062209 (2020).

\bibitem{naikoo2020} J. Naikoo, S. Dutta, and S. Banerjee, Phys. Rev. A \textbf{99}, 042128 (2019).

\bibitem{passos2019} M. H. M. Passos, P. C.  Obando, W. F. Balthazar, F. M.  Paula, J. A. O. Huguenin and M. S. Sarandy,  Opt. Lett. \textbf{44}, 2478 (2019).

\bibitem{rana} S. Rana, P. Parashar, and M. Lewenstein, Phys. Rev. A \textbf{93}, 012110 (2016).

\bibitem{chsh} J. F. Clauser, M. A. Horne, A. Shimony, and R. A. Holt, Phys. Rev. Lett. \textbf{23}, 880 (1969).

\bibitem{PD-mem} C. Macchiavello and G.P.  Massimo, Phys. Rev. A \textbf{65}, 050301 (2002).

\bibitem{nmd2018} U. Shrikant, R. Srikanth, and S. Banerjee, Phys. Rev. A  \textbf{98}, 032328 (2018).

\bibitem{rtn1} N. P. Kumar, S. Banerjee, R. Srikanth, V. Jagadish,  and  F. Petruccione,  Open Syst. Inf. Dyn. \textbf{25}, 1850014 (2018).

\bibitem{pln} S. Utagi, R.  Srikanth, and S. Banerjee, Sci. Rep. \textbf{10}, 15049 (2020).

\bibitem{baan} N. B. An and J. Kim, Phys. Rev. A  \textbf{79}, 022303 (2009).

\bibitem{AD-mem} Y. Yeo and A. Skeen, Phys. Rev. A \textbf{67}, 064301 (2003).

\bibitem{nmad1} B. Bellomo, R. L. Franco,  and  G. Compagno,  Phys. Rev. Lett. \textbf{99}, 160502 (2007).

\bibitem{nmad2} K. Thapliyal, A. Pathak,  and S. Banerjee,  Quantum Inf. Process.  \textbf{16}, 115 (2017).

\bibitem{tp1} C. H. Bennett, G. Brassard, C. Crépeau, R. Jozsa, A Peres, and W. K. Wootters,  Phys. Rev. Lett. \textbf{70}, 1895 (1993).

\bibitem{tp2} D. Bouwmeester, J. W. Pan, K. Mattle, M. Eibl, H. Weinfurter, and A. Zeilinger, Nature  \textbf{390}, 575 (1997).

\bibitem{pop} S. Popescu, Phys. Rev. Lett. \textbf{72}, 797 (1994).

\bibitem{werner2} C. H. Bennett, G. Brassard, S. Popescu, B. Schumacher, J. A. Smolin, and W. K. Wootters, Phys. Rev. Lett. \textbf{76}, 722 (1996).

\bibitem{f1max} R. Horodecki, M. Horodecki, and P. Horodecki, Phys. Lett. A \textbf{222}, 21 (1996)

\bibitem{hu_tele} M. L. Hu, Quantum Inf. Process.  \textbf{12}, 229 (2013).

\end{thebibliography}
\end{document}